\documentclass[a4paper,twoside,prd,nofootinbib,preprintnumbers,twocolumn]{revtex4}
\usepackage{amssymb,amsmath,bm,natbib}
\usepackage{color}
\usepackage{slashed}
\usepackage{graphics}
\usepackage{graphicx}
\usepackage[utf8]{inputenc}
\usepackage[caption=false]{subfig}
\usepackage{hyperref}
\usepackage{url}
\usepackage{dsfont}
\usepackage{float} 
\usepackage{cancel}

\newcommand{\eq}[1]{Eq.~\eqref{#1}}

\begin{document}
\preprint{CERN-TH-2021-036, PSI-PR-21-03, ZU-TH 11/21}

\title{Correlating Non-Resonant Di-Electron Searches at the LHC to \\
	the Cabibbo-Angle Anomaly and Lepton Flavour Universality Violation}

\author{Andreas Crivellin}
\email{andreas.crivellin@cern.ch}
\affiliation{CERN Theory Division, CH--1211 Geneva 23, Switzerland}
\affiliation{Physik-Institut, Universit\"at Z\"urich, Winterthurerstrasse 190, CH--8057 Z\"urich, Switzerland}
\affiliation{Paul Scherrer Institut, CH--5232 Villigen PSI, Switzerland}

\author{Claudio Andrea Manzari}
\email{claudioandrea.manzari@physik.uzh.ch}
\affiliation{Physik-Institut, Universit\"at Z\"urich, Winterthurerstrasse 190, CH--8057 Z\"urich, Switzerland}
\affiliation{Paul Scherrer Institut, CH--5232 Villigen PSI, Switzerland}

\author{Marc Montull}
\email{marc.montull@psi.ch}
\affiliation{Physik-Institut, Universit\"at Z\"urich, Winterthurerstrasse 190, CH--8057 Z\"urich, Switzerland}
\affiliation{Paul Scherrer Institut, CH--5232 Villigen PSI, Switzerland}

\begin{abstract}
In addition to the existing strong indications for lepton flavour university violation (LFUV) in low energy precision experiments, CMS recently released an analysis of non-resonant di-lepton pairs which could constitute the first sign of LFUV in high-energy LHC searches. In this article we show that the Cabibbo angle anomaly, an (apparent) violation of first row and column CKM unitarity with $\approx3\,\sigma$ significance, and the CMS result can be correlated and commonly explained in a model independent way by the operator $[Q_{\ell q}^{(3)}]_{1111} = (\bar{\ell}_1\gamma^{\mu}\sigma^I\ell_1)(\bar{q}_1\gamma_{\mu}\sigma^Iq_1)$. This is possible without violating the bounds from the non-resonant di-lepton search of ATLAS (which interestingly also observed slightly more events than expected in the electron channel) nor from $R(\pi)=\pi \to\mu\nu/\pi \to e \nu$. We find a combined preference for the new physics hypothesis of $4.5\,\sigma$ and predict $1.0004<R(\pi)<1.0009$ (95\%~CL) which can be tested in the near future with the forthcoming results of the PEN experiment.
\end{abstract}
\maketitle

\newpage
\section{Introduction}

The Standard Model (SM) of particle physics has been very successfully tested with great precision in the last decades. Nonetheless, it is clear that the SM cannot be the ultimate fundamental theory of physics. For example, it has to be extended to account for Dark Matter and Neutrino masses, but neither the scale nor the concrete nature of the additional particles necessary to explain these observations is unambiguously established. Fortunately, in the flavour sector intriguing (indirect) hints for physics beyond the SM at the (multi) TeV scale have been collected in the last years. In particular, lepton flavour universality has been tested extensively, unveiling intriguing signs of beyond the SM physics in $b\to s\ell^+\ell^-$~\cite{Aaij:2014pli,Aaij:2014ora,Aaij:2015esa,Aaij:2015oid,Khachatryan:2015isa,ATLAS:2017dlm,CMS:2017ivg,Aaij:2017vbb} data, $b\to c\tau\nu$~\cite{Lees:2012xj,Lees:2013uzd,Aaij:2015yra,Aaij:2017deq,Aaij:2017uff,Abdesselam:2019dgh} transitions, and the anomalous magnetic moment of the muon~\cite{Bennett:2006fi} with a significance of $>\!5\,\sigma$~\cite{Capdevila:2017bsm,Altmannshofer:2017yso,DAmico:2017mtc,Ciuchini:2017mik,Hiller:2017bzc,Geng:2017svp,Hurth:2017hxg,Alok:2017sui,Alguero:2019ptt,Aebischer:2019mlg,Ciuchini:2019usw,Ciuchini:2020gvn}, $>\!3\,\sigma$~\cite{Amhis:2016xyh,Murgui:2019czp,Shi:2019gxi,Blanke:2019qrx,Kumbhakar:2019avh} and $3.7\,\sigma$~\cite{Aoyama:2020ynm}, respectively. Furthermore, the Cabibbo-Angle anomaly (CAA), a deficit in first row (and first column) CKM unitarity with a significance of $3\,\sigma$~\cite{Belfatto:2019swo,Grossman:2019bzp,Seng:2020wjq,Coutinho:2019aiy}, can also be explained within the framework of LFUV beyond the SM~\cite{Coutinho:2019aiy,Crivellin:2020lzu}. However, until recently, no hints of LFUV at high energy searches at the LHC had emerged.

This changed when CMS recently reported the results of a first test of LFUV in non-resonant di-lepton searches by measuring the di-muon to di-electron ratio~\cite{Sirunyan:2021khd}, observing an excess in the electron channel. Interestingly, also ATLAS found slightly more electron events than expected in the signal region in the search for quark-lepton contact interactions~\cite{Aad:2020otl} and HERA reported more electron events than expected as well~\cite{Abramowicz:2019uti}, even though the bounds are not competitive with the ones from the LHC. Clearly, this surplus of events can very well be just a statistical fluctuation, and CMS actually states that no significant tension with the SM prediction is observed. However, it appears in the bins with high invariant mass of the electron pair, as expected in case of heavy NP that can be parameterized in terms of effective 2-quark-2-lepton operators. Therefore, an investigation of the implications of this excess is very interesting and in this article we want to show that the CAA can be model-independently correlated to non-resonant di-electron searches. In fact, assuming that the CAA is explained by a direct NP contribution to beta decays, a signal in di-electron production is even predicted, whose size turns out to agrees with the data reported by CMS and is compatible with the ATLAS bounds. In addition, such an explanation predicts an observable effect in $R(\pi) = \pi\to\mu\nu/\pi\to e\nu$ (defined at the amplitude level) which perfectly agrees with the current data and can be soon tested by the forthcoming results from the PEN experiment. 

In the next section we define our setup and discuss the different observables. We combine these observables and show the results of the global analysis in Sec.~\ref{results} before we conclude and give an outlook in Sec.~\ref{conclusions}.

\begin{figure*}[t!]
	\centering
	\includegraphics[width=0.75\textwidth]{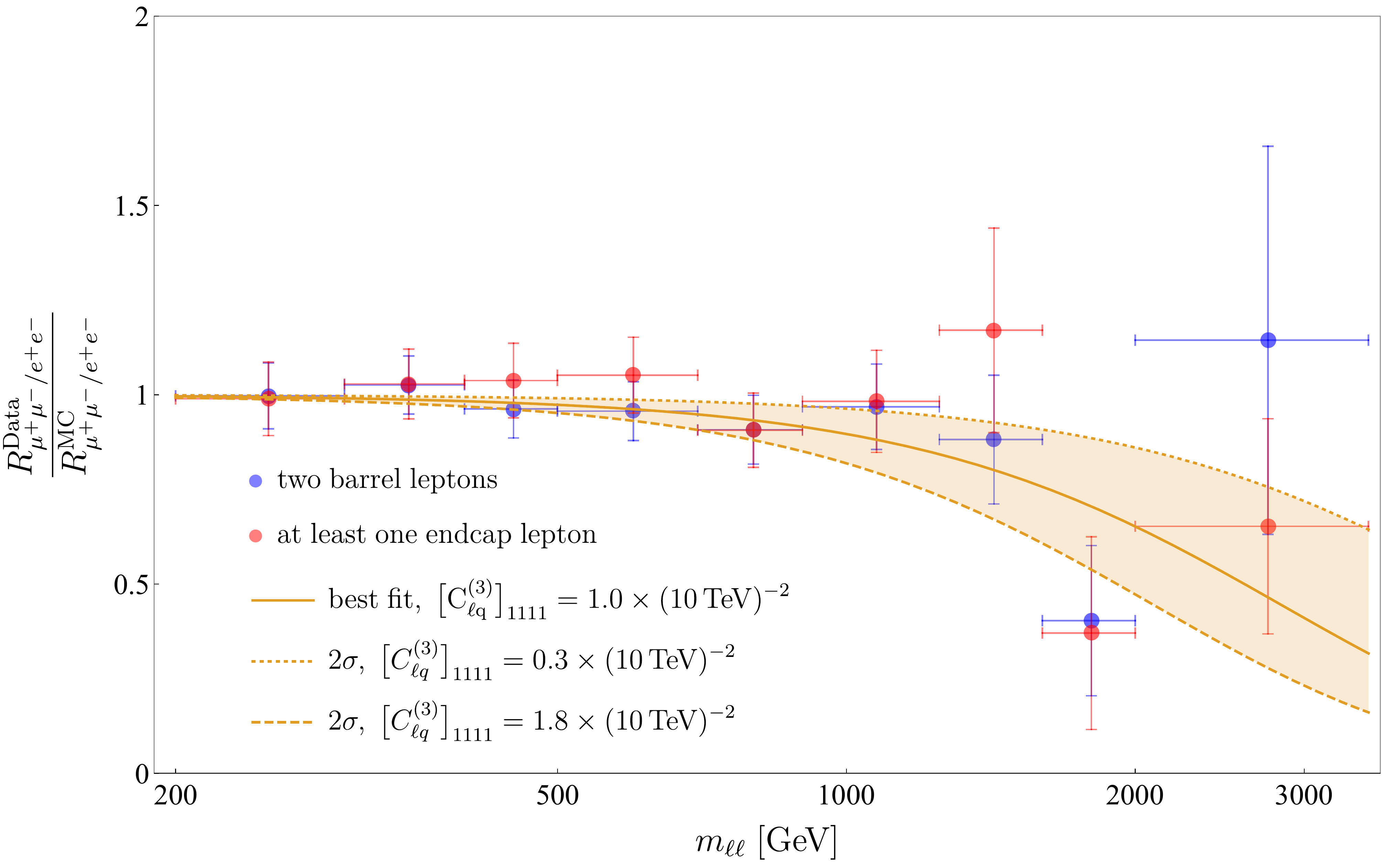} 
	\hspace{10mm}
	\caption{ Ratio of the differential di-lepton production cross section in the di-muon and di-electron channels as a function of the lepton pair invariant mass for events with two barrel leptons (blue) and at least one lepton in the endcaps (red)~\cite{Sirunyan:2021khd}. The error bars include both statistical and systematic uncertainties. In dark orange the predicted number of events for the best fit point of $[C_{\ell q}^{(3)}]_{1111}$ and the $2 \sigma$ region.
	\label{CMSPlot}}
\end{figure*}

\section{Setup and Observables}
\label{setup}

As outlined in the introduction, the CAA is the deficit found in first row and first column CKM matrix unitarity~\cite{Belfatto:2019swo,Grossman:2019bzp,Seng:2020wjq,Coutinho:2019aiy}
The tension significantly depends on the radiative corrections to superallowed $\beta$ decays~\cite{Marciano:2005ec,Seng:2018yzq,Seng:2018qru,Gorchtein:2018fxl,Czarnecki:2019mwq,Seng:2020wjq,Hayen:2020cxh,Hardy:2020qwl} and on the treatment of the $K_{\ell 2}$ and $K_{\ell 3}$ decays~\cite{Moulson:2017ive} as well as the constraints from $\tau$ decays~\cite{Amhis:2019ckw} (see Ref.~\cite{Crivellin:2020lzu} for more details). However, a significance of around $3\sigma$ should give a realistic estimate of the current situation. For definiteness we use the result of Ref.~\cite{Zyla:2020zbs} 
\begin{align}
\big|V_{ud}\big|^2+\big|V_{us}\big|^2+\big|V_{ub}\big|^2
&= 0.9985(5)\,,\label{firstrow}\\
\big|V_{ud}\big|^2+\big|V_{cd}\big|^2+\big|V_{td}\big|^2 &= 0.9970(18)\,.
\label{1throw}
\end{align}
Note that even though the deficit in the first column CKM unitarity is
less significant than the one of the first row, it suggests that, if the deficits were due to BSM effects they would likely be related to $\beta$ decays and therefore to $V_{ud}$. For the numerical analysis, we will only use the relation for the first row due to its higher precision.  

There are several possibilities to account for the CAA~\cite{Crivellin:2021njn}. For instance, via modified $W$-quark couplings~\cite{Belfatto:2019swo,Belfatto:2021jhf}, a modified $W-\mu\nu_\mu$ coupling~\cite{Coutinho:2019aiy,Bryman:2019bjg,Crivellin:2020lzu,Capdevila:2020rrl,Endo:2020tkb,Crivellin:2020ebi,Kirk:2020wdk,Alok:2020jod}, a tree-level contribution to the muon decay~\cite{Belfatto:2019swo,Crivellin:2020oup,Crivellin:2020klg} or by a tree-level effect in beta decays~\cite{Crivellin:2021egp}. Since we aim at connecting the CAA to the CMS measurement we will focus on the latter possibility in which case only one operator, in the basis of Ref.~\cite{Grzadkowski:2010es}, 
\begin{align}
	[Q_{\ell q}^{(3)}]_{1111} = (\bar{\ell}_1\gamma^{\mu}\sigma^I\ell_1)(\bar{q}_1\gamma_{\mu}\sigma^Iq_1)\,,
\end{align}
(where $\sigma^I$ are the Pauli matrices) is capable of explaining the CAA without violating other bounds, in particular those set by $R(\pi)=\pi\to\mu\nu/\pi\to e\nu$~\cite{Crivellin:2021njn}. This four-fermion operator generates effects in the neutral and charged-current processes after EW symmetry breaking via the Lagrangian
\begin{align}
	\begin{split}
	\mathcal{L} = \mathcal{L}_{\rm SM} &+  [C_{\ell q}^{(3)}]_{1111}\bigg[(\bar{d}\gamma^{\mu}P_L d -\bar{u}\gamma^{\mu}P_L u)\bar{e}\gamma_{\mu}P_L e\\
	&\!\!\!+ (\bar{u}\gamma^{\mu}P_L u-\bar{d}\gamma^{\mu}P_L d)\bar{\nu}\gamma_{\mu}P_L \nu\\
	&\!\!\!+ 2\left(\bar{d}\gamma^{\mu}P_Lu\bar{\nu}\gamma_{\mu}P_L e + \bar{u}\gamma^{\mu}P_Ld\bar{e}\gamma_{\mu}P_L \nu\right) \bigg]\,,
	\end{split}
\end{align}
where we omitted CKM matrix elements. Note, that in principle, after CKM rotations, we could get effects in processes like $K\to\mu\nu/K\to e\nu$ or $K\to\pi\nu\nu$. However, these bounds can be avoided by assuming that $[C_{\ell q}^{(3)}]_{11kl}$ is aligned to the down basis or flavour universality in the quark sector (i.e. $[C_{\ell q}^{(3)}]_{11kl}=\delta_{kl} [C_{\ell q}^{(3)}]_{1111}$).
A non-zero Wilson coefficient of the operator $[Q_{\ell q}^{(3)}]_{1111}$ modifies the CKM unitarity relations. Using \eq{firstrow} we find that the best fit point for the Wilson coefficient is
\begin{align}
[C_{\ell q}^{(3)}]_{1111}= 1.22(4)/(10 \, \text{TeV})^2\,.
\end{align} 

$Q_{\ell q}^{(3)}]_{1111}$ also contributes to non-resonant di-electron production at the LHC, which is tailored to search for heavy NP that is above the direct production reach~\cite{Eichten:1984eu,Eichten:1983hw} and therefore can be parametrized in terms of an effective Lagrangian. The latest di-lepton results from ATLAS and CMS are presented in Ref.~\cite{Aad:2020otl} and Ref.~\cite{Sirunyan:2021khd}, respectively. 

Even though ATLAS does not claim any tension with the SM prediction, they observe $19$ $e^+ e^-$ events in the signal region for the case of constructive interference compared to an expectation of $12.4\pm 1.9$ events\footnote{Here the error includes only the systematic but the statistical uncertainty.}. This leads to a limit that is weaker than expected but which agrees with the SM expectation within $2\,\sigma$ once the statistical uncertainty is included. We have to recast the ATLAS limit on the Wilson coefficient since their bound is obtained for operators that have equal coupling to up and down quarks while $[Q_{\ell q}^{(3)}]_{1111}$ leads to an relative minus sign between them\footnote{Here we corrected Eq.~(1) of Ref.~\cite{Aad:2020otl} by a missing factor 2. We thank the ATLAS collaboration for confirming this typo.}. Recasting the case with constructive interference with the SM we find that
\begin{equation}
[C_{\ell q}^{(3)}]_{1111}\lessapprox 1.4/(10\,{\rm TeV})^2\,,
\end{equation}
at $95\%\,$CL.  We obtained this limit by using the differential parton-anti-parton luminosities~\cite{Campbell:2006wx} which for partons $i,j=u,d,s,c,b$ are given by
\begin{equation}
\frac{d \mathcal{L}_{i \bar j}}{d \hat{s}}=\frac{1}{s} \int_{\tau}^{1} \frac{d x}{x} f_{i}(x, \sqrt{\hat{s}}) f_{j}\left(\frac{\tau}{x}, \sqrt{\hat{s}}\right)+(i \leftrightarrow \bar j) \,,
\end{equation}
where $s$ ($\hat{s}$) is the beam (parton) center of mass energy and $\tau = \hat{s}/s$. The integrated cross section over $\hat{s}$ can then be computed as
\begin{equation}
\sigma=\sum_{i, \bar j} \int\left(\frac{d \hat{s}}{\hat{s}}\right)\left(\frac{d \mathcal{L}_{i \bar j}}{d \hat{s}}\right)\left(\hat{s} \hat{\sigma}_{i \bar j}\right) \,,
\end{equation}
where $\hat{\sigma}_{ab}$ is the partonic cross section which is typically a function of $\hat s$. For the numerical evaluation we use the PDF set NNPDF23LO, also employed e.g. by ATLAS analysis to generate the signal DY process~\cite{Aad:2020otl} with the help of the \texttt{Mathematica} package \texttt{ManeParse}~\cite{Clark:2016jgm}. We then computed at tree-level the cross section for our NP model, normalized to the SM one, as a function of the invariant mass of the lepton pair, integrated this over the invariant mass of the lepton pairs within the signal region $m_{\bar{\ell} \ell} \in [2.2,6] \, \text{TeV}$ and compared it to the limits obtained from ATLAS.

CMS observed a slight excess in the di-electron cross section at high invariant lepton mass and computed the double ratio 
\begin{equation}
R^{\rm Data}_{\mu^+\mu^-/e^+e^-}/R^{\rm MC}_{\mu^+\mu^-/e^+e^-}\,,
\end{equation}
in order to reduce the uncertainties~\cite{Greljo:2017vvb}. This means that they provide the relative signal strength for muons vs electrons, $R^{\rm Data}_{\mu^+\mu^-/e^+e^-}$, divided to the SM expectation obtained from Monte Carlo simulations, $R^{\rm MC}_{\mu^+\mu^-/e^+e^-}$. Importantly, in this procedure the first bin is normalized to one in order to obtain the relative sensitivity to electrons and muons. Taking this into account, we find that the best fit value for the Wilson coefficient is 
\begin{equation}
[C_{\ell q}^{(3)}]_{1111}\approx1.0 /(10\,{\rm TeV})^2\,,
\end{equation}
with $\Delta \chi^2 \equiv \chi^2 - \chi^2_{SM} \approx-10$ and $0.3 /(10\,{\rm TeV})^2\lessapprox [C_{\ell q}^{(3)}]_{1111}\lessapprox1.8 /(10\,{\rm TeV})^2$ at 95\%~CL as shown in Fig.~\ref{CMSPlot}. Here we followed the same approach outlined above for the ATLAS analysis and computed the ratio of the cross section in our model with respect to the tree-level SM one. Since the cross section is already dominated in the SM by left-handed amplitudes, we can assume that the changes in the angular distributions, affecting the CMS analysis, are small and can be safely neglected.

\begin{figure*}[t!]
	\centering
	\includegraphics[width=0.7\textwidth]{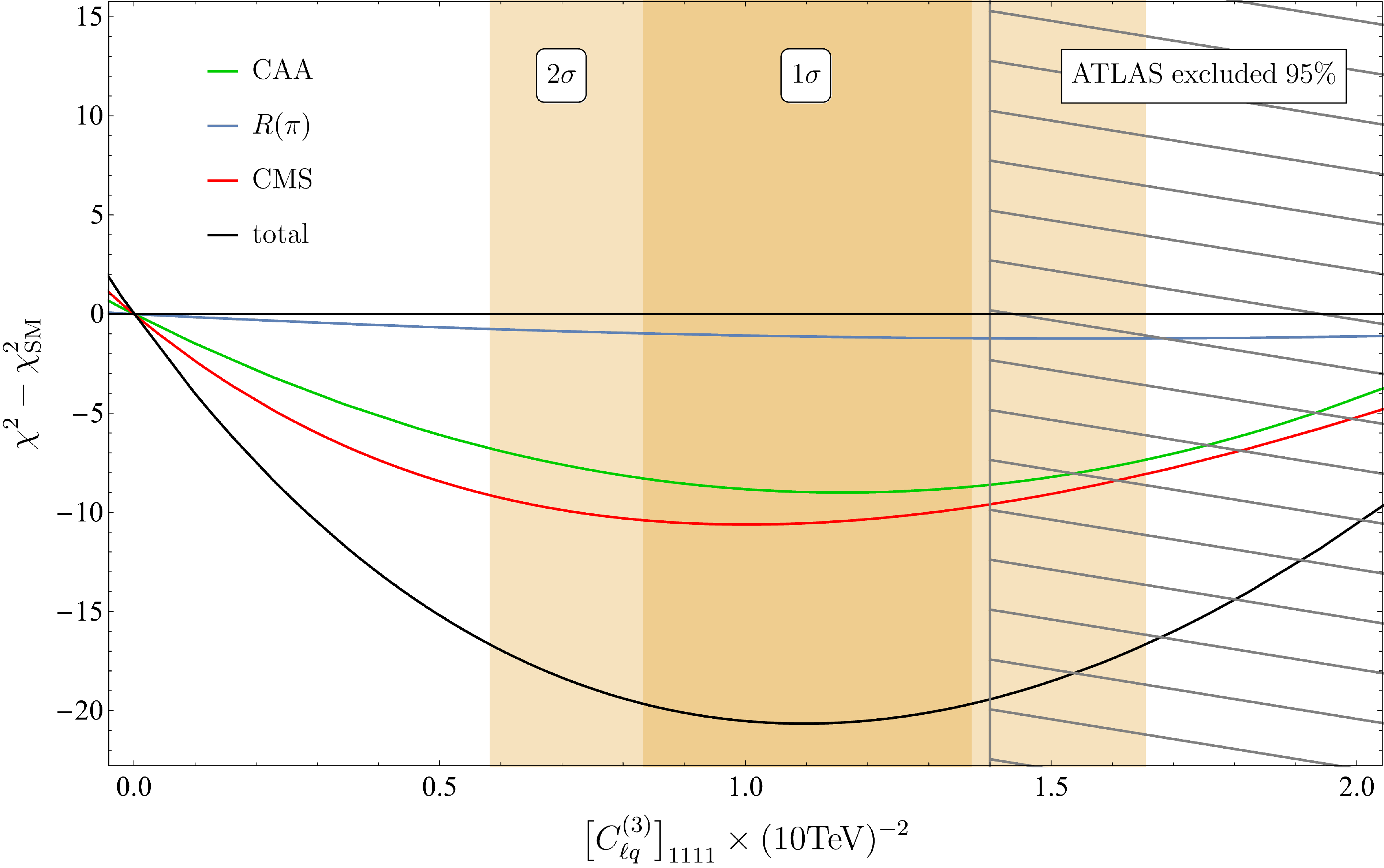} 
	\hspace{10mm}
	\caption{
		$\Delta \chi^2 = \chi^2 - \chi^2_{SM}$ as a function of the Wilson coefficient $[C_{\ell q}^{(3)}]_{1111}$ for the fits including only $R(\pi)$ (blue), the CAA (green), the CMS analysis of di-lepton pairs  (red) and the combination of them (black) with the dark orange region showing the $1 \sigma$ and $2 \sigma$ regions of the combined fit. The hatched region is excluded at 95\% CL from the non-resonant di-lepton search of ATLAS (not included in the $\chi^2$ function). The best fit point for the combined fit (black) is at $[C_{\ell q}^{(3)}]_{1111}\approx1.1/(10\,{\rm TeV})^2$ where $\chi^2-\chi^2_{SM} \approx - 20$.  \label{chi2plot}	}
\end{figure*}

Finally we note that our operator also gives a LFUV effect which can be tested by charged pion decay. Defining the ratio $ R(\pi) = \pi\to\mu\nu/\pi\to e\nu$ at the amplitude level we have
\begin{align}
	\begin{split}
	 R\left(\pi\right) \simeq 1 + 0.0006 \,  \,  [C_{\ell q}^{(3)}]_{1111} \times (10 \, \text{TeV})^2 \,,
	\end{split}
	\label{LFUratios}
\end{align}
which has to be compared with the experimental measurement~\cite{Czapek:1993kc,Britton:1992pg,Bryman:1982em,Cirigliano:2007xi,Aguilar-Arevalo:2015cdf,Tanabashi:2018oca}
\begin{align}
\begin{split}
&R\left(\pi\right) = 1.0010 \pm 0.0009 \,.
\end{split}
\end{align}
Note that even though there is no deviation from the SM prediction, the $1\sigma$ interval is in perfect agreement with the expectations from the CAA and CMS data.

\section{Combined Analysis and Results}\label{results}

Let us now perform the combined analysis of the observables discussed in the last section. In Fig.~\ref{chi2plot} we show the $\Delta\chi^2=\chi^2-\chi^2_{\rm SM}$ for the CMS measurement of LFUV in non-resonant di-lepton searches, the CAA and $R(\pi)$ separately, as well as the 95\%~CL exclusion region from ATLAS. We compute the total $\Delta\chi^2$ function which has a minimum for $[C_{\ell q}^{(3)}]_{1111}\approx1.1/(10\,{\rm TeV})^2$ of $\approx-20$, corresponding to a pull of $\approx 4.5\,\sigma$ with respect to the SM. Note that this minimum is well compatible with the 95\% CL exclusion limit of ATLAS, which however cuts partially the $2\sigma$ region preferred by $R(\pi)$, CMS and the CAA. Treating the ATLAS exclusion as a hard cut, we therefore find that at $95\%$ CL
\begin{equation}
0.6/(10{\rm TeV})^2\lessapprox [C_{\ell q}^{(3)}]_{1111} \lessapprox 1.4/(10{\rm TeV})^2 \,.
\end{equation}
This interval corresponds to a prediction for $R(\pi)$ of
\begin{equation}
1.0004\lessapprox R(\pi)\lessapprox 1.0009\,,
\end{equation}
at the 95\% CL This prediction can be tested by the forthcoming results of PEN~\cite{Glaser:2018aat} and PiENu~\cite{Mischke:2018qmv} experiments which anticipate in the near future an improvement by more than a factor $3$. Furthermore, the proposed PiENuXe aims at an order of magnitude improvement in sensitivity compared to the current experimental result.

\section{Conclusions and Outlook}\label{conclusions}

While convincing hints for the violation of LFU have been accumulating within  recent years by low energy precision experiments, corresponding signals in high energy searches at the LHC have not been found in the past. However, recently, a  non-resonant di-lepton analysis of CMS showed an excess in electrons compared to muons. Even though this can very well just be a statistical fluctuation, it is interesting that the surplus of events is the bins with the highest invariant mass of the lepton pair (as expected if it is due to a heavy NP contribution) and both ATLAS and HERA also observe more electron events than expected in similar analyses. 

Furthermore, the Cabibbo angle anomaly (the deficit in first row and first column CKM unitarity at the $3\,\sigma$ level) can be addressed via a tree-level NP contribution to beta decays. Such an explanation must involve the operator $[Q_{\ell q}^{(3)}]_{1111}$ of the SMEFT which leads to an effect in non-resonant di-electron searches at the LHC with a signal strength compatible with the one found by CMS. In fact, we find that a combined explanation of CMS data and the Cabibbo angle anomaly with this operator can improve the total SM $\chi^2$ by more than 20, corresponding to a pull of $\approx 4.5\,\sigma$, while respecting the bounds from the corresponding ATLAS search.

Our scenario can be tested by analyses of the forthcoming LHC Run 3 data and, to an even better degree, at the high luminosity LHC or the LHeC~\cite{Agostini:2020fmq}. Moreover improvements in the determinations of $V_{us}$, $V_{ud}$ and also $V_{cd}$, both on the theoretical and experimental side (see Refs.~\cite{Crivellin:2020lzu,Crivellin:2021njn} for a more detailed discussion), will scrutinize the deficit in first row and column CKM unitarity. In addition, we predict that in our setup $1.0004\lessapprox R(\pi)\lessapprox 1.0009$ at 95\% CL which can soon be tested by the PEN and PiENu experiments. Furthermore, as we performed an EFT analysis, this opens up new possibilities in model building where e.g. $C_{\ell q}^{(3)}$ could be generated by a vector triplet, allowing for correlations with $b\to s\ell^+\ell^-$ data~\cite{Capdevila:2020rrl}.
\medskip

{\it Acknowledgments} --- {\small 
We thank Emanuele Bagnaschi, Noam Tal Hod, Massimiliano Procura, Oliver Fischer and Daniel Lechner for useful discussions.	This work is supported by a Professorship Grant (PP00P2\_176884) of the Swiss National Science Foundation. A.C. thanks CERN for the support via the scientific associate program.	 
}

\bibliography{bibliography}

\end{document}